\documentclass[fleqn,usenatbib]{mnras}

\usepackage{newtxtext,newtxmath}

\usepackage[T1]{fontenc}

\DeclareRobustCommand{\VAN}[3]{#2}
\let\VANthebibliography\thebibliography
\def\thebibliography{\DeclareRobustCommand{\VAN}[3]{##3}\VANthebibliography}

\usepackage{orcidlink}
\usepackage{graphicx}	
\usepackage{amsmath}	
\usepackage[normalem]{ulem} 
\usepackage{comment}	
\usepackage{soul}               

\newcommand{\eg}{\textit{e.g.}}

\defcitealias{lee1994}{LTC94}

\title[COWS: A cosmic filament finder]{COWS: A filament finder for Hessian cosmic web identifiers}

\author[S. Pfeifer et al.]{Simon Pfeifer,$^{1}$\thanks{E-mail: spfeifer@aip.de}
Noam I. Libeskind\orcidlink{0000-0002-2769-9507}$^{1,2}$, Yehuda Hoffman$^3$, Wojciech A. Hellwing$^4$, \newauthor Maciej Bilicki$^4$, \& Krishna Naidoo$^{4,5}$
\\
$^{1}$Leibniz-Institut für Astrophysik Potsdam, An der Sternwarte 16, D-14482 Potsdam, Germany\\
$^{2}$Univ Lyon, Univ Claude Bernard Lyon 1, CNRS, IP2I Lyon / IN2P3, IMR 5822, F-69622, France\\
$^{3}$Racah Institute of Physics, Hebrew University, Jerusalem 91904, Israel\\
$^{4}$Center for Theoretical Physics, Polish Academy of Sciences, Al. Lotników 32/46, 02-668 Warsaw, Poland\\
$^{5}$Department of Physics \& Astronomy, University College London, Gower Stree, London, WC1E 6BT, UK\\}
\date{Accepted XXX. Received YYY; in original form ZZZ}

\pubyear{2021}

\begin{document}
\label{firstpage}
\pagerange{\pageref{firstpage}--\pageref{lastpage}}
\maketitle
\setstcolor{red}

\begin{abstract}
The large scale galaxy and matter distribution is often described by means of the cosmic web made up of voids, sheets, filaments and knots. Many different recipes exist for identifying this cosmic web. Here we focus on a sub-class of cosmic web identifiers, based on the analysis of the Hessian matrix, and proposed a method, called COsmic Web Skeleton (COWS), of separating a set of filaments cells into an ensemble of individual discreet filaments. Specifically, a thinning algorithm is applied to velocity shear tensor based cosmic web (V-web) to identify the spine of the filaments. This results in a set of filaments with well defined end-point and length. It is confirmed that these sit at local density ridges and align with the appropriate direction defined by the underlying velocity field. The radial density profile of these curved cylindrical filaments, as well as the distribution of their lengths is also examined. The robustness of all results is checked against resolution and the V-web threshold. The code for the COWS method has been made publicly available. 
\end{abstract}

\begin{keywords}
keyword1 -- keyword2 -- keyword3
\end{keywords}



\section{Introduction}
The  Cosmic Web \citep{bond1996} is one of the most intriguing and striking patterns found in nature, rendering its analysis and characterization 
far from trivial. Quantifying the large scale distribution of matter beyond $N$-point correlation functions calls for the examination of cosmic fields. Indeed, the intricate multi-scale cosmic network is observable using one of a number of (correlated) physical quantities, among them the density field, the shear field, the tidal field, the potential field or the velocity field \citep{hahn2007,hoffman2012}.

The roots of quantifying large-scale structure using fields can be traced to the seminal work of Zel'dovich and his famous ``pancake'' (or blini) metaphor for structure formation \citep{zeldovich1970}. Accordingly, the universe is endowed with small (aspherical) density perturbations which grow via gravitational instability to form large structures observable in the galaxy distribution. These structures are formed by matter first collapsing along one axis creating pancakes (or cosmic sheets). The subsequent phase of gravitational instability occurs along a perpendicular axis causing the Zel'dovich pancake to then roll up into a filament\footnote{The use of the word ``roll'' here is not flippant as it is believed that it is during this second stage when angular momentum is generated  by tidal torques \citep{doroshkevich1970,white1984}.}. Such curvi-linear gravitational structures are not stable and ultimately collapse  forming aspherical blobs (also termed cosmic knots or clusters). 
This approach to structure formation led to the Soviet school's ``top-down'' picture described above (\eg \citealt{arnold1982,klypin1983}), and is slightly divergent from the ``hierarchical'', or ``bottom-up'', paradigm which pervades the way we think about (small scale) structure formation in $\Lambda$CDM. But on large (linear or quasi-linear) scales, the cosmic web's hierarchy appears ``dimensional'' - voids are conceived as aspherical bubbles (3D), sheets, like the supergalactic plane, are flattened structures (2D), cosmic filaments are (curvi-)linear (1D) and clusters or knots are gravitational sinks (0D), corresponding to sites of peculiar velocity flow line convergence. In fact the ordering of these web types -- voids, sheets, filaments, knots -- is a natural outcome of examining the differential compression or expansion of matter encapsulated by the tidal or velocity shear tensor; simply put this ordering arises naturally from counting along how many axes matter is locally expanding: 3, 2, 1 or 0, respectively.

In the case of a potential flow, the velocity shear tensor is the Hessian of the velocity potential. A diagonalization  of the tensor at each point in space and an examination of the eigenvalues allows for the quantification of all space into one of the four cosmic web types, according to the hierarchy described above \citep{hoffman2012}. Numerous authors have examined the velocity shear field, correlating its principle axis with physical propertied of haloes and galaxies including the direction of infalling material, the orientation of planes of satellites, and galaxy spin \citep[\eg][]{forero-romero2014,libeskind2015}.

When segmenting the universe into dynamically defined structures using the tidal or velocity shear tensor, most studies have simply identified which volumes of the universe share the same web classification \citep{hahn2007b,forero-romero2009,cautun2013,libeskind2013}. Practically, the velocity or density field is put onto a three dimensional grid and each cell in this mesh is given a classification based on the values of the eigenvalues of the diagonalized tensor. While this approach can study the global properties, e.g. of filaments, as a whole, more detailed studies require the identification of individual cosmic web objects. 

Identifying filaments in both the galaxy distribution and simulations is important as a means to understand various physical processes related to galaxy formation. As quasi-linear objects they are interesting because matter has collapsed along all three axes but only along one of these is the flow still laminal - along the other two there has been shell crossing. They are quasi-linear because viewed along one axis the flow look linear, while viewed along the other two its clearly non-linear.

Filaments constitute the immediate environment within which ``field'' galaxies form.  The properties of galaxies have been extensively shown to depend on environment \citep{dressler1980,avila-reese2005,blanton2005,gao2005,maulbetsch2007,forero-romero2011} and to properties of dark matter haloes \citep[\eg][]{hahn2007,Hellwing2021}. In addition, filaments are likely intimately related to galaxy spin \citep{codis2012,tempel2013,wang2017,ganeshaiah2019} and therefore their morphology. 

There are already a number of filament finders in the literature including the highly successful Bisous model \citep{stoica2014,tempel2016} as well as Skeleton \citep{sousbie2008}, \texttt{DisPerSE} \citep{sousbie2011} and \texttt{NEXUS} \citep{cautun2013} which introduced adaptive scaling in the context of filament finding (see \citet{libeskind2018} for a summary of filament and cosmic web finders in general). \citet{cautun2014} introduced a method to compress filaments (and walls) towards their central axis (plane). The linear, compressed filament network can then be separated into individual filaments based on a criterion of the rate of change of filament orientation (i.e. filaments that bend with acute angles below a threshold are separated). This work follows a similar approach to filament finding. Here, we presents a method to identify the spines of filaments from Hessian-based Cosmic Web classifiers, specifically the V-web.

This paper is organised as follows: The simulations and the relevant data extracted from them are explained in Section~\ref{sec:data}. Section~\ref{sec:method} presents the complete method of generating the filament catalogue from the V-web. Section~\ref{sec:results} presents results aimed at validating the algorithm, as well as a selection of interesting statistics such as the radial density profiles of filaments. Section~\ref{sec:conclusion} summarises and discusses the results.

\section{Data}
\label{sec:data}
The filament finding method is applied to a cosmological $N$-body simulation of periodic box size of 400 comoving Mpc/$h$ on a side which contains $1024^3$ particles run with a modified version of \texttt{Gadget-3}  (last described in \citealt{gadget}). The simulation is part of the \texttt{BAHAMAS} suite of simulations \citep{mccarthy2017,pfeifer2020}. Initial conditions were generated using a modified version of \texttt{N-GenIC}\footnote{\url{https://github.com/sbird/S-GenIC}} with transfer functions at a starting redshift of $z=127$ computed by \texttt{CAMB}\footnote{\url{http://camb.info/}}\citep{lewis1999}. The cosmology is almost equivalent to the \textit{Planck} 2015 best-fit cosmology \citep{planck2015} with $h=0.69$, $\Omega_{\rm m}=0.294$, $\Omega_{\Lambda}=0.706$, $n_s=0.974$ and $\sigma_{8}=0.802$.

To calculate the shear tensor from the distribution of the 1024$^3$ DM particles the velocity field is constructed on a regular grid by means of  ``Clouds-in-Cell'' (CIC) interpolation. The largest size (resolution) of the grid is indirectly determined by the particle resolution. The empirical rule of $(N/4)^3$ cells for $N$ particles is used in order to ensure few if any cells are empty. Therefore, a grid with 256$^3$ cells is used given the particle number mentioned above. The CIC of the velocities returns the momentum field and therefore is divided by the number of particles in their respective cells to account for this.

The discretised velocity field is then smoothed with a Gaussian kernel to remove any spurious artefacts due to the artificial cartesian grid. A kernel size of down to 1 cell width, in this case (400 Mpc/$h$)/256 = 1.56 Mpc/$h$, may be employed. Once the CIC of the velocity field has been smoothed, the shear tensor may be computed. The Gaussian smoothing and velocity shear calculation is performed in Fourier space using Fast Fourier Transform. Using a lower resolution grid with the smallest smoothing kernel returns very similar results to simply using a higher resolution grid with the same size smoothing kernel. Therefore the smoothing can be thought of as the effective resolution. 

 \begin{figure*}
    \centering
    \includegraphics[width=0.9\textwidth]{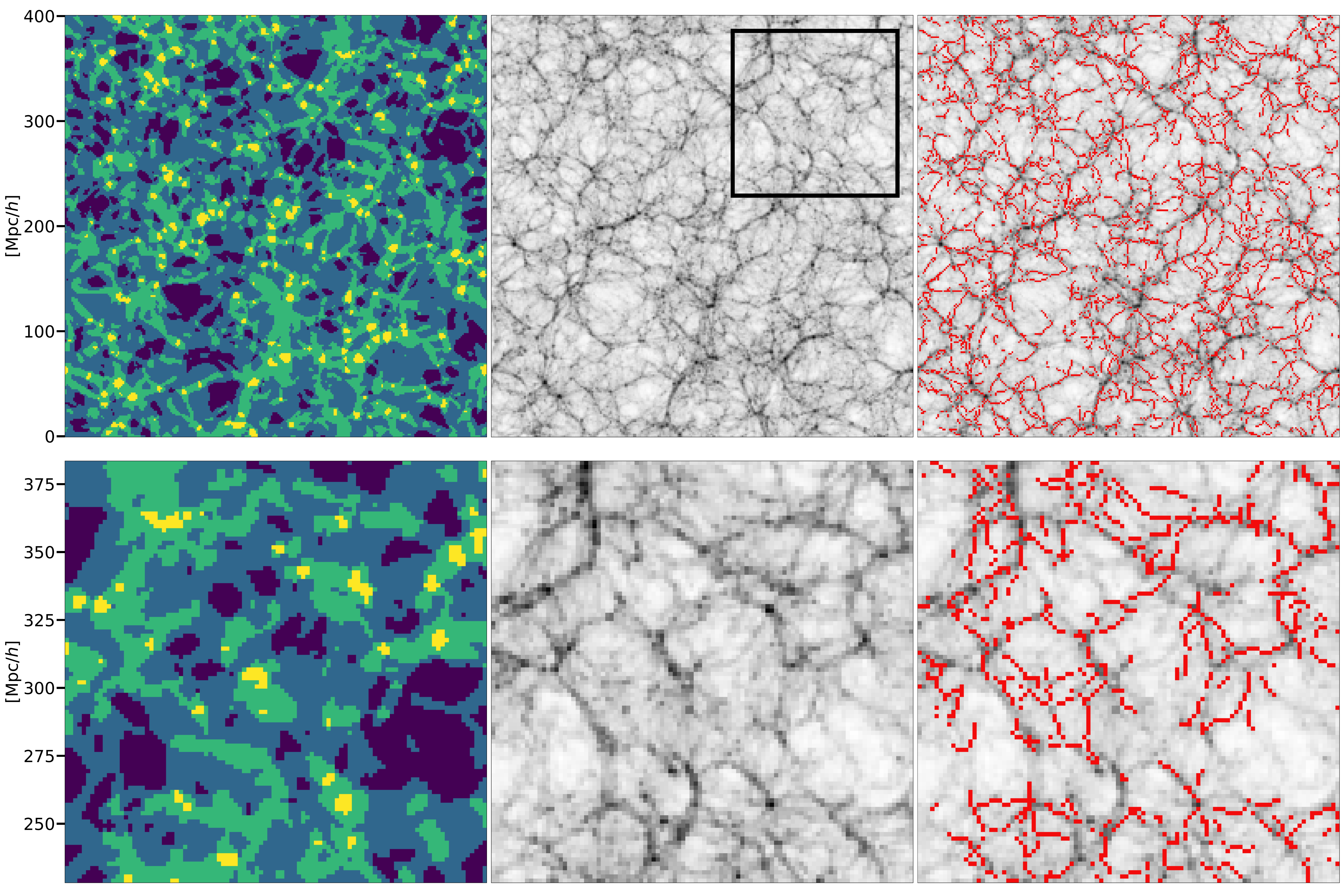}
    \caption{Slices of the DM-only simulation with thickness of $\approx$15 Mpc/$h$. Particles have been discretized onto a regular grid using the "Cloud-in-Cell" scheme using 256 cells along each axis. The size of the Gaussian smoothing kernel for the V-web calculation is set to 1.56 Mpc/$h$ (one cell width) and the V-web threshold, $\lambda_{\rm th}$, is set to 0. \textit{Left}: The V-web output (Section \ref{sec:vweb}) at the centre of the simulation slice. Voids, sheets, filament and knots are shown in order from darkest (purple) to brightest (yellow). \textit{Middle}: The logarithmic projected density of the dark matter. Darker areas indicate higher density. \textit{Right}: The individual filaments extracted from the V-web are overplotted in red over the density. \textit{Bottom}: A zoom into the region indicated by the black square in the top middle panel. Each column shows the same type of data as the panel above.}
    \label{fig:quadplot}
\end{figure*}

\section{Method}
\label{sec:method}
The COsmic Web Skeleton (COWS) method is briefly described here with more detailed descriptions of each aspect given below. We start by computing the V-web on a regular grid using the velocities of all particles from a dark matter-only cosmological simulation. As we work exclusively with the V-web, we will refer to the velocity shear tensor as simply the shear tensor henceforth. The V-web assigns each grid cell to one of the four main cosmic web types: knots, filaments, sheets or voids. Sheets and voids are discarded by setting their corresponding cells to a background value. The grid now contains an interconnected structure of filaments and knots that extends throughout the simulation box. We identify the medial axes, or skeleton, of this structure by applying an iterative thinning procedure called medial axis thinning. The skeleton is made up of connected cells that follow the approximate geometric centre of the V-web filament structure and have widths of one cell. Each cell of the skeleton can then be labeled based on its local connectivity. Cells with 2 neighbours, basic cells, and 1 neighbour, end points, are kept, and all other cells are set to the background value. This results in a set of skeleton edges, analogous to filament spines, each with well defined end points and length which can be seen on the right in Fig.~\ref{fig:quadplot}. Below, the method of calculating the V-web is described in Section~\ref{sec:vweb}, the medial axis thinning algorithm is described in Section~\ref{sec:mat} and the final filament identification is described in Section~\ref{sec:skelclass}. The implementation of the COWS method has been made publicly available\footnote{The code is available at \href{https://github.com/SimonPfeifer/cows}{https://github.com/SimonPfeifer/cows}.}. It uses modified versions of methods from the scikit-image library \citep{scikit-image}.
 
\subsection{V-web}
\label{sec:vweb}
The V-web method is described below but for a more detailed description see \citet{hoffman2012} and \citet{libeskind2012,libeskind2013}. The V-web is used to classify the cosmic web on a grid using the shear tensor. The shear tensor is defined at each grid cell as 
\begin{equation}
    \Sigma_{\alpha\beta} = - \frac{1}{2H_0}\left(\frac{\partial  v_{\alpha}}{\partial  r_{\beta}} + \frac{\partial  v_{\beta}}{\partial  r_{\alpha}}\right),
    \label{equ:vweb}
\end{equation}
where $\alpha,\beta=x,y,z$ and $H_{0}$ is the Hubble constant. Note that that $H_0$ is introduced  to make the shear tensor tensor dimensionless.

The velocity shear is calculated for each cell and then diagonalized to calculate the eigenvalues ($\lambda_1$, $\lambda_2$, $\lambda_3$) and eigenvectors ($\hat{e}_{1}$, $\hat{e}_{2}$, $\hat{e}_{3}$), following the convention of ordering the eigenvalue and corresponding eigenvectors such that $\lambda_1$>$\lambda_2$>$\lambda_3$. 

A minus sign is added to the right-hand side of the above expression such that positive eigenvalues corresponds to convergence along that eigenvector. Matter is said to be collapsing along $\hat{e}_{i}$ if the corresponding eigenvalue, $\lambda_{i}$, is greater than a threshold values, $\lambda_{\rm th}$. Conversely matter is said to be expanding along $\hat{e}_{i}$ if the corresponding eigenvalue $\lambda_{i}$ is less than $\lambda_{\rm th}$. The V-web is defined by counting the number of eigenvalues above $\lambda_{\rm th}$, each grid cell may be classified: 0, 1, 2, 3 for voids, sheets, filaments and knots, respectively. The value of $\lambda_{\rm th}$ has typically been chosen such that the V-web return a visual impression of the cosmic web and usually lies within the range [0,1] \citep{hoffman2012}. 
 
 A V-web distribution of a simulation slice is shown in the left column of Fig.~\ref{fig:quadplot}. The colours show voids, sheets, filaments and knots from dark to bright, respectively. The corresponding logarithmic projected (summed along the line of sight) density slice is shown in the middle column of Fig.~\ref{fig:quadplot} where dark regions indicated higher density. Comparing these two panels visually, one can see that the structures present in the density are captured by the V-web. Dark, high density regions are identified as knots and their connecting regions as filaments.

\begin{figure*}
    \centering
    \includegraphics[width=\textwidth]{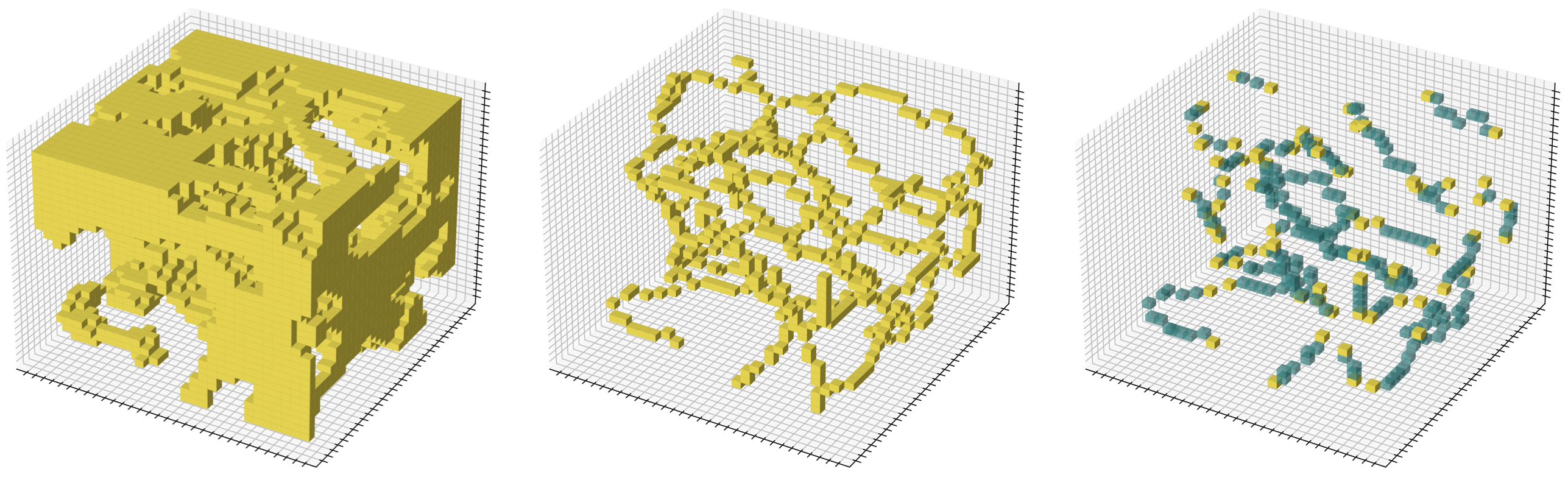}
    \caption{3D distributions of the stages of the filament finding method in a 30x30x30 cube. \textit{Left}: The filament and knots cells from the V-web that make up the input for the medial axis thinning. \textit{Middle}: The skeleton output of the medial axis thinning. The main features and shape of the input data are preserved (see bottom left of cube for example). \textit{Right}: The extracted filaments from the skeleton. Each filament has its endpoints marked in solid yellow.}
    \label{fig:3dplot}
\end{figure*}

\subsection{Medial axis thinning}
\label{sec:mat}
In this section we explain the first step in how individual filaments are constructed from the V-web classification. The method outline below follows that described in \citet{lee1994} (LTC94 henceforth) and the reader is referred to that paper for a thorough description. Medial axis thinning (MAT) is employed largely in the field of computer graphics to reduce large amounts of data in 2D images or 3D models while preserving geometric features of the data. A medial axis can be thought of as a skeleton of a geometric structure, and we will refer to it as such throughout. The method presented in \citetalias{lee1994} takes a 3D grid of binary data and reduces it by iteratively removing border cells while satisfying topological and geometric constraints. Below, a brief explanation of the specific constraints which dictate whether a specific cell can be removed are explained.

Given that the data is required to be binary (i.e. a cell is either in the structure or not), we define the set of cells with a value of 1 as $S$ and the set of cells with a value of 0 as $\Bar{S}$. In practice, $S$ are all cells identified as knots and filament by the V-web, and $\Bar{S}$ are all remaining cells (sheets and voids). A cell, $v$, in $S$ is called a border point if one or more of it 6-neighbours are in $\Bar{S}$, where the 6-neighbours, $N_6(v)$, are all the cells that it shares a face with.

The first condition (C1 in \citetalias{lee1994}) requires the topology of the data to be invariant if a given border point were to be removed. The topology of a geometry can be defined by the Euler characteristic  which is invariant for manifolds with the same topology. Instead of calculating the Euler characteristic of the entire geometry, it is possible to only consider a predefined neighbourhood around a given cell, such as the 26-neighbours, $N_{26}(v)$, which are all the neighbouring cells in a 3x3x3 cube, i.e. all directly adjacent, diagonal and corner neighbours. To simplify the problem further, the $N_{26}(v)$ space is split into eight overlapping 2x2x2 cubes referred to as octants. The advantage to this approach is that an octant has only 256 configuration which allows for the Euler statistic for all configuration to be precalculated and stored. \citetalias{lee1994} derive an equation for the change in the Euler characteristic of an octant and calculate this value for all 256 possible configurations, given the removal of the same point in the octant, and provide the corresponding look-up table. In practice, one needs to identify the configuration of the eight octants around a cell and, for each octant, look-up the change in the Euler characteristic (labeled $8 \delta G_{26}$ in \citetalias{lee1994}) using the octant's configuration. If the sum of the changes to the Euler characteristic for all eight octants does not equal zero, the cell cannot be removed as the topology would not be conserved.

The second condition (C2) checks if a given cell is classed as a `simple' point. \citetalias{lee1994} review many definitions of a simple point, where their chosen definition has been proven to be a complete set. Removing simple points conserves many topological properties. To test if a cell is a simple point, \citetalias{lee1994} show that the number of connected objects across $N_{26}(v)$ must remain constant if the cell is removed. Two cells are connected if a path of cells across their $N_{26}(v)$ exists between them and these cells are part of the same set $S$. A set of points which are connected in such a way are defined as a connected object. \citetalias{lee1994} propose a labeling algorithm using an octree data structure that returns the number of connected objects in $N_{26}(v)$ and provide pseudo code that was used to implement this method.
The next condition (C4) checks if a given cell is an end point of an arc. \citetalias{lee1994} claim that the simple definition that an end point of an arc only has one neighbour is sufficient to extract the medial axis if all other conditions are satisfied. 

The last condition (C3) addresses the practical problem that arises by simultaneously removing all identified simple border points. Checking the condition outlined above for all cells can be easily performed in parallel, resulting in a set of simple border points proposed for removal. However, some proposed simple border points will be neighbours and removing one of these will affect the outcome of the conditions of the other. \citetalias{lee1994} explicitly show configurations for which the simultaneous removal of proposed simple border points does not conserve the topology. To resolve this issue, \citetalias{lee1994} propose a sequential re-checking procedure. For each proposed simple border point, remove it and then check the number of connected objects in $N_{26}(v)$ using the algorithm proposed for C2. If the number of connected objects is larger than one, the proposed simple border point cannot be removed. This condition is only required if all simple border points are identified first and then removed. If the conditions C1, C2 and C4 are checked sequentially and each simple border point is immediately removed, C3 is not needed.

\subsection{Skeleton classification and filament identification}
\label{sec:skelclass}
The MAT method returns a skeleton structure that follows the approximate centre of the V-web structure, shown in the middle panel of Fig.~\ref{fig:3dplot}. The next step is to remove any contaminants or undesired features and identify individual filaments. Individual filaments should have well defined end points and an unambiguous, single main branch, i.e. contain no forks.

\citetalias{lee1994} propose a method for labeling the cells that make up the skeleton based on the number of neighbours a cell has in $N_{26}(v)$. Cells with 1, 2, 3, and 4 neighbours are called end points, regular cells, T-junctions and X-junctions, respectively. Regular cells make up the majority of the skeletons. End points and regular cells are easily identified. However, \citetalias{lee1994} present configurations in which multiple adjacent cells are classified as either T-junctions and X-junctions when a single junction cell is preferred. A proposed solution to this misclassification is to force all neighbouring cells in $N_{26}(v)$ of a junction to be regular cells. This method will contaminate the set of regular cells with cells that have more than 2 neighbours and join filaments that should be separate in the method described below. We therefore do not adopt this method.

\begin{figure*}
    \centering
    \includegraphics[width=0.9\textwidth]{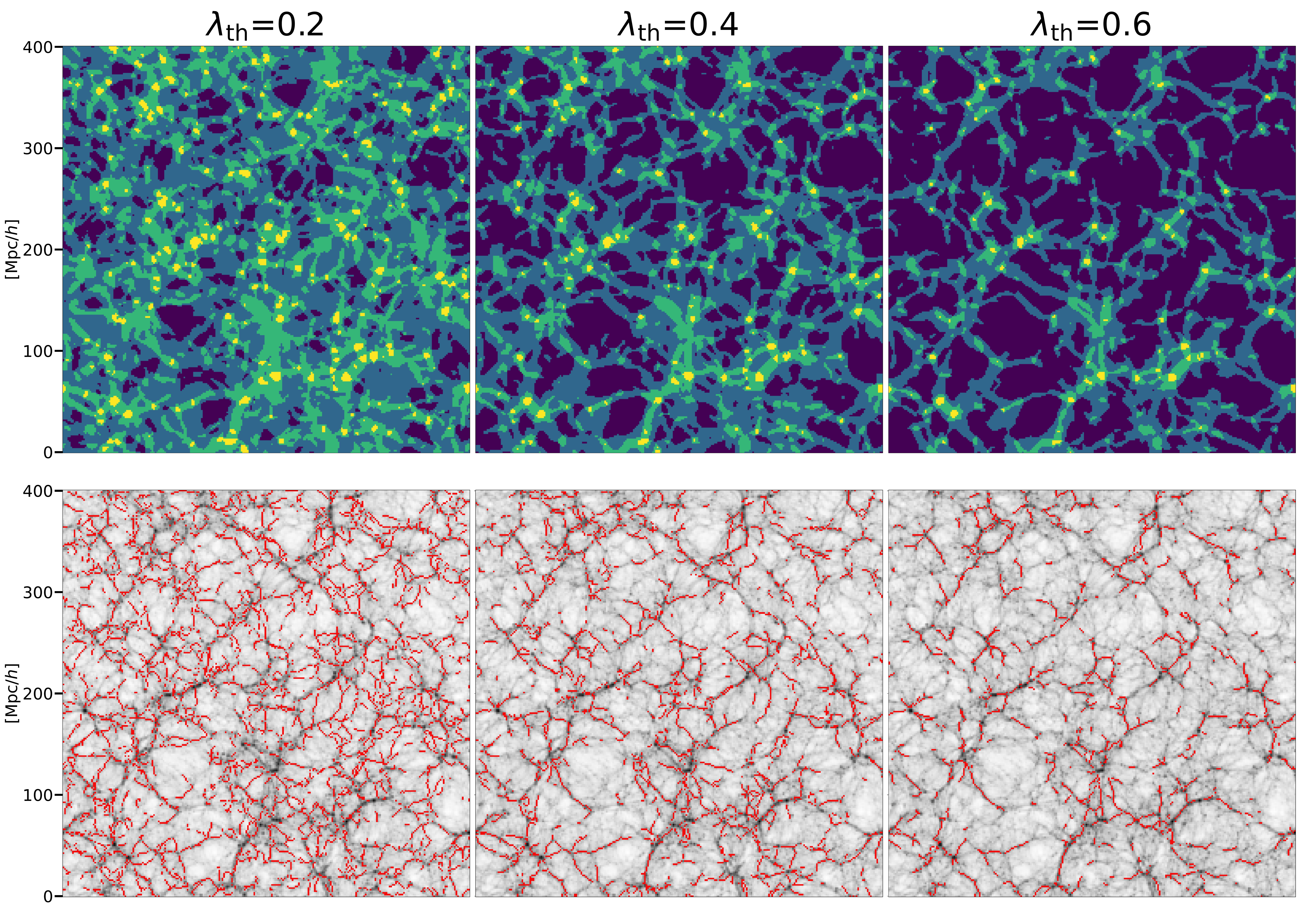}
    \caption{Panels showing the effect of varying the V-web threshold, $\lambda_{\rm th}$. Each column refers to a different value of $\lambda_{\rm th}$, increasing from left to right. The size of the Gaussian smoothing kernel is 1.56 Mpc/$h$ for all panels. \textit{Top row:} V-web slices of the simulation, same as the top right panel of Fig.~\ref{fig:quadplot}. From dark to bright, the colours indicate voids, sheets, filaments and knots. \textit{Bottom row:} The resulting filaments in red, plotted over the logarithmic density of DM in a ~15 Mpc/$h$ slice, where darker areas indicate higher density.}
    \label{fig:triplot_thetas}
\end{figure*}

MAT conserves topological features which include cavities in the V-web structure. Cavities are a set of cells $\Bar{S}$ that are completely surrounded by a set of cells $S$. These cavities become spherical, hollow features in the skeleton that are undesired and need to be removed. The member cells of these cavities tend to have many neighbours in $N_{26}(v)$.

To classify the skeleton, all cells are labeled with a number that is equal to the number of neighbours in $N_{26}(v)$. All cells with more than 2 neighbours are removed. It is important to first label all cells and then remove the appropriate ones as checking each cell and immediately removing it  will affect the classification of neighbouring cells which will lead to inconsistent results depending on the order in which the cells are checked. This removes all T-junctions, X-junctions and cavity features. A second pass is performed, labeling all cells with the number of neighbours they have. This 
returns a separated skeleton containing only regular cells and end points.

Single filaments can be identified by joining cells recursively in $N_{26}(v)$, starting at a cell labeled as an end point, until a second end point is reached or, equally, until no more cells can be joined. Each filament is made up of two end points joined by a single path of regular cells with width of one cell and every cell is a member of only a single filament. These can be seen in the right panel of Fig.~\ref{fig:3dplot}.

\section{Results}
\label{sec:results}

The following properties of the filaments are extracted from the target simulation: position, orientation, the distribution of their length and density profiles perpendicular to the filaments spines. These are used, in part, to validate the COWS method and study the effects of varying V-web threshold, $\lambda_{\rm th}$, and resolution.

\subsection{Filament orientation and position}
Of course, the simplest method to check the accuracy of the filament finder is to inspect the density structure of the simulation and the corresponding filaments by eye. Fig.~\ref{fig:quadplot} shows the projected (summed along the line of sight) DM density in a $\approx$15 Mpc/$h$ thick slice of the simulation and the corresponding filaments overplotted in red. The filaments identified by COWS trace the density ridges that connect large local density maxima such as knots, and trace the majority of the visible filaments very well. It is important to note that not all filament-like structures in the density field projection are in fact cosmic filament. For example, a 2D projected slice through a cosmic sheet could also appear filamentary. These visual comparisons are therefore used to gain a general impression of the performance of the filament finder.

The method for extracting filaments from the V-web has no free parameters and will therefore always return the same filament catalogue for a given V-web distribution. However, the V-web itself has a free parameter, the threshold $\lambda_{\rm th}$, which directly affects the cosmic web type that is assigned to a cell. A second parameter that can be varied is the number of cells used for the regular grid, or alternatively, the Gaussian smoothing performed during the V-web method. Both affect the effective resolution of the binned simulation data. Therefore, it is of interest to investigate how the filament finder behaves when changing $\lambda_{\rm th}$ and the size of the Gaussian smoothing kernel. 

\subsubsection{Varying V-web threshold}
\label{sec:varythreshold}
The dependence of the filament finding on the V-web threshold, $\lambda_{\rm th}$, is considered first. Fig~\ref{fig:triplot_thetas} shows V-web slices (top) and the corresponding density field with overplotted COWS filaments in red (bottom) for different values of $\lambda_{\rm th}$. Raising the value of $\lambda_{\rm th}$ effectively shifts the V-web cell classification away from knots and towards voids. This has a thinning effect on knots and filaments while sheets and voids assimilated new cells. This can be seen in Fig~\ref{fig:triplot_thetas} where the left panel, corresponding to $\lambda_{\rm th} =0.2$, contains only a few voids while the right panel, corresponding to $\lambda_{\rm th} =0.6$, is almost completely void cells. Inspecting the resulting filaments, the first thing to note is that lower values of $\lambda_{\rm th}$ allows the filaments to trace the lower density, small-scale structures whereas higher values trace only the main, high density branches. Increasing $\lambda_{\rm th}$ also results in fewer cavity artefacts in the skeleton. The extracted filaments tend to be longer and their endpoints lie closer to the centre of adjacent knots when $\lambda_{\rm th}$ is large. By changing the value of $\lambda_{\rm th}$ from 0 to 0.6 in steps of 0.2, the filament finder identifies 39011, 22028, 12010 and 6823 individual filaments, effectively halving the number of filaments for each increase of 0.2 in $\lambda_{\rm th}$.

The first test of our filament finding algorithm is to assess the orientation of the filament spines. To check the orientation of the filament sample, a direction needs to be assigned to each filament cell that is identified. Consider a filament cell $i$ that has two neighbours, namely cell $i-1$ and $i+1$. Two distance vectors can be constructed, namely $d_{i-1,i}=r_{i-1}-r_{i}$ and $\hat{d}_{i,i+1}=r_{i}-r_{i+1}$, where $r$ refers to the cell position. A direction unit vector, $\hat{d}$, of a given filament cell is then simply
\begin{equation}
    \hat{d} = \frac{d_{i-1,i}+d_{i,i+1}}{|d_{i-1,i}+d_{i, i+1}|} = \frac{r_{i-1}-r_{i+1}}{|r_{i-1}-r_{i+1}|},
    \label{equ:fildirection}
\end{equation}
where the direction should be considered an axis rather than a vector since the order in which neighbours are considered is arbitrary. For the end points in filaments which have only a single neighbour, either of the terms on the right-hand side of Equation~\ref{equ:fildirection} can be set to the position of the filament cell resulting in a normalised vector pointing towards/from the single neighbour. Using Equation~\ref{equ:fildirection}, each filament cell is assigned a direction.

\begin{figure}
    \centering
    \includegraphics[width=\columnwidth]{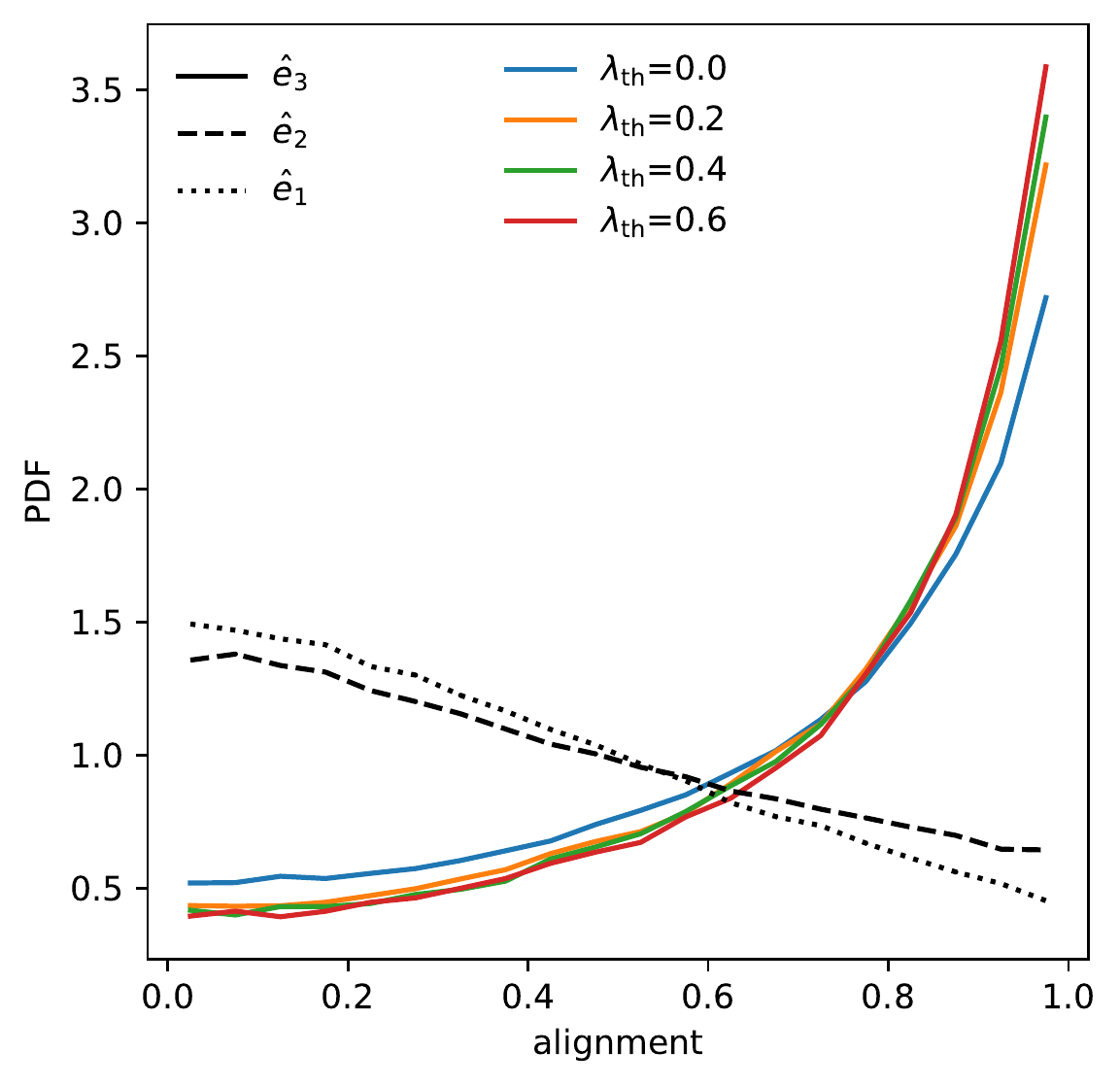}
    \caption{Probability density function (PDF) of the alignment between filament direction and the $\hat{e}_3$ eigenvector of the V-web. Filament direction is calculated using Equation~\ref{equ:fildirection}. Alignment is calculated as the absolute value of the dot product between the two vectors. Colours indicate different values of the V-web threshold, $\lambda_{\rm th}$. For reference, the alignment for $\hat{e}_1$ and $\hat{e}_2$ are plotted for $\lambda_{\rm th}=0$ and indicated by different linestyles.}
    \label{fig:filalignment_lambda}
\end{figure}

\begin{figure}
    \centering
    \includegraphics[width=\columnwidth]{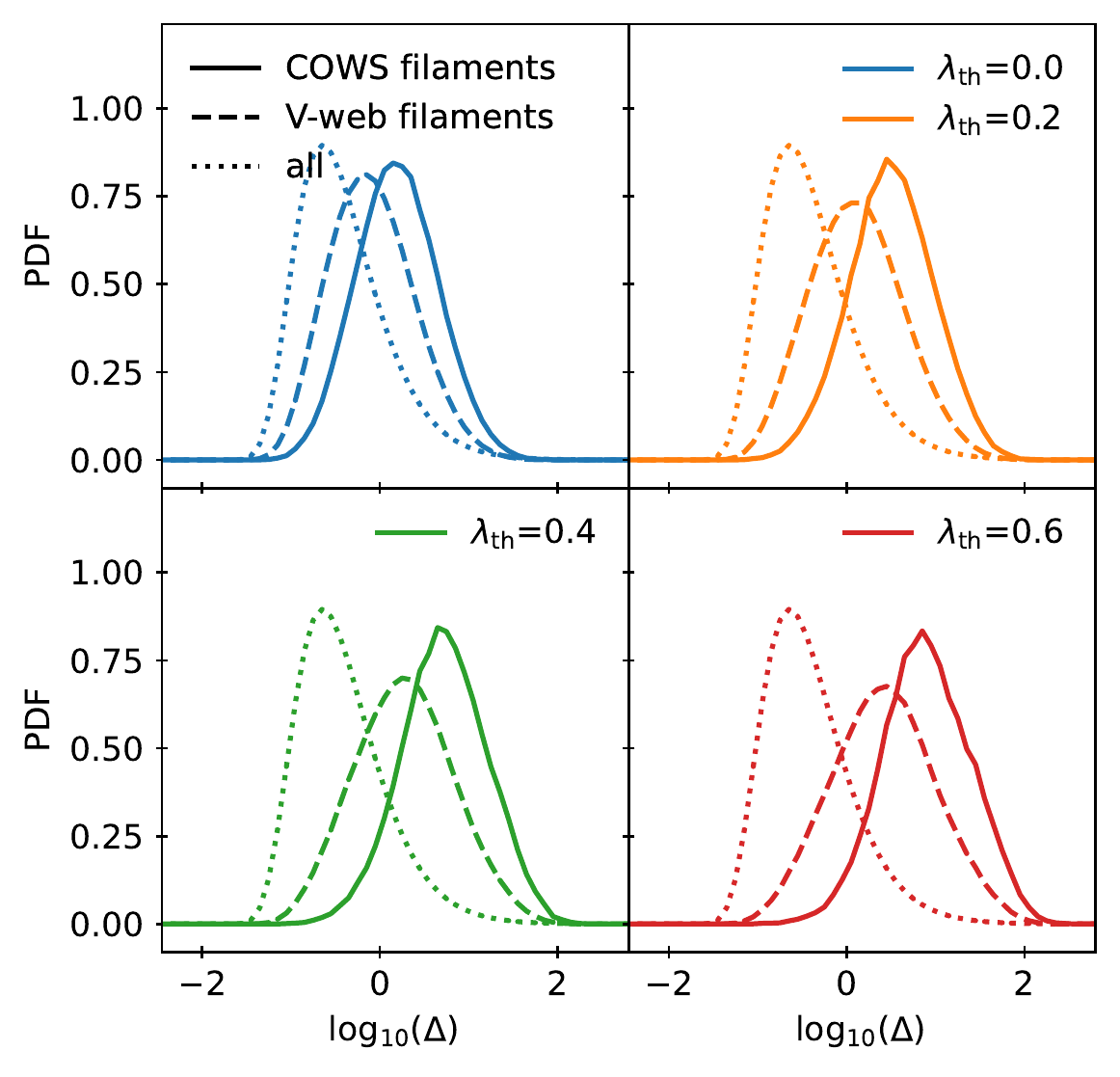}
    \caption{Probability density function (PDF) of the logarithmic normalised density for the identified filaments (solid), V-web filament cells (dashed) and all cells in the simulation (dotted). Different colours indicate different values of the V-web threshold, $\lambda_{\rm th}$.}
    \label{fig:filoverdensity_lambda}
\end{figure}

Here we check the orientation of the filaments. The local shear tensor endows three principal directions for each point in space. To check how the filaments are orientated with respect to the principal directions of the underlying shear tensor field, the dot product of the direction $\hat{d}$ of each filament cell and the eigenvectors of the underlying local shear tensor is calculated and  presented in Fig~\ref{fig:filalignment_lambda}. The alignment is calculated by taking the absolute value of the dot product between the filament direction and each shear tensor eigenvector, since neither of these axes have directions. A set of parallel axes will have an alignment of one and a set of perpendicular axes would give zero. A random distribution of axis directions would result in a flat PDF with a constant value of one. It is important to note here that the filament direction is an inferred quantity. In other words, the
filament direction does not know about the direction of the underlying velocity shear field and thus a comparison with the velocity shear eigenvectors is fair test of the filament finder. 

\begin{figure*}
    \centering
    \includegraphics[width=0.9\textwidth]{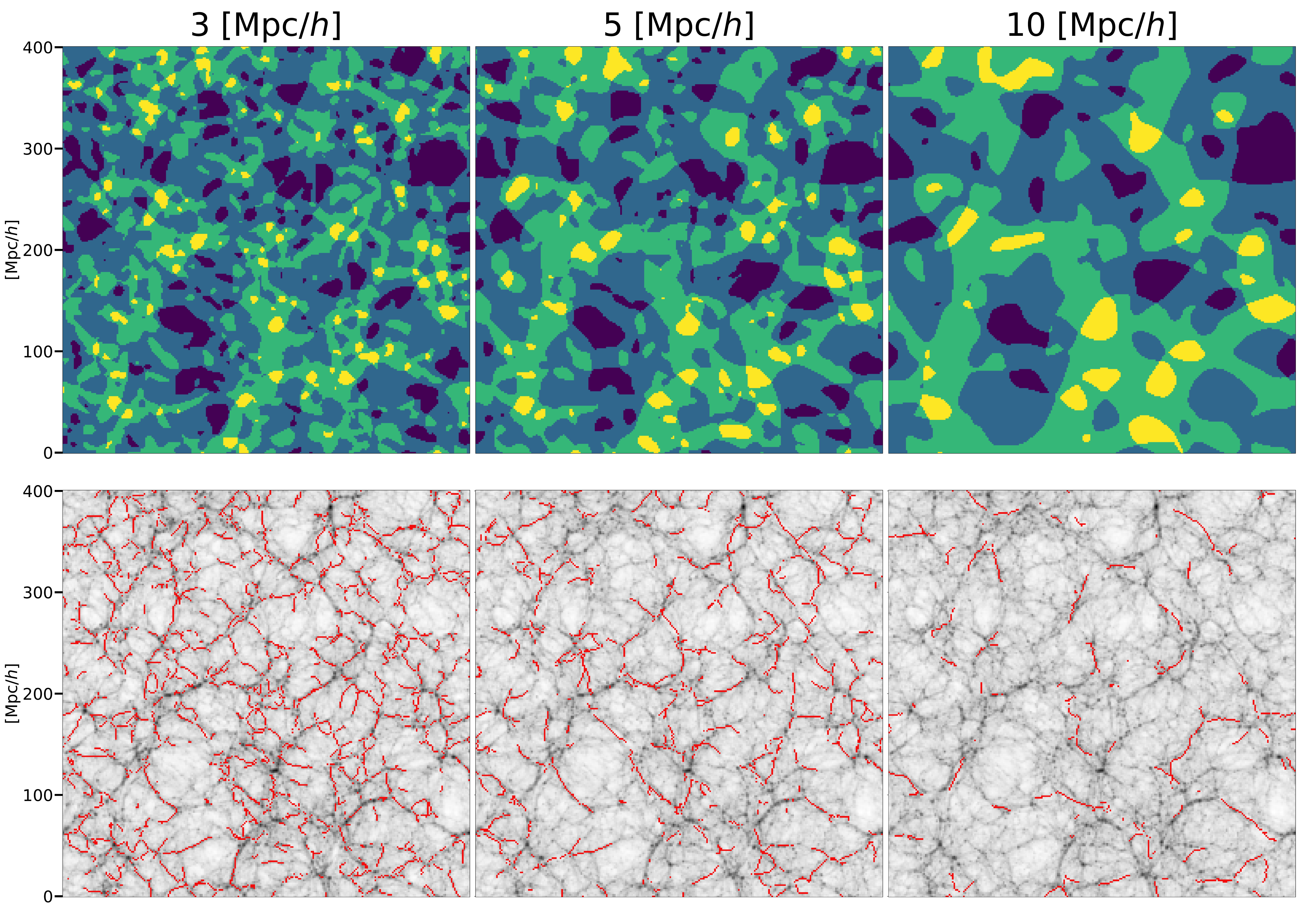}
    \caption{Panels showing the effect of varying the size of the Gaussian smoothing kernel in the V-web calculation. Each column refers to a different kernel size, increasing from left to right. The value of $\lambda_{\rm th}$ is zero for all panels. Each row shows the same as Fig.~\ref{fig:triplot_thetas} except that the density field has been smoothed with their respective Gaussian kernel and the thickness of the slice has increase in proportion to the size of the kernel for a fairer comparison.}
    \label{fig:triplot_smoothscale}
\end{figure*}

Fig~\ref{fig:filalignment_lambda} shows that the filaments preferentially align with the $\hat{e}_3$ eigenvector and tend to be perpendicular to $\hat{e}_1$ and $\hat{e}_2$. The $\hat{e}_3$ eigenvector points along the axis of expansion or, alternatively, the axis of least collapse. The alignment is robust to changes in the V-web threshold where increasing $\lambda_{\rm th}$ fractionally increases the alignment with $\hat{e}_3$.

The position, specifically the density at the filament spine positions, is examined next. By construction, the identified filaments are a subset of all the cells identified as filaments from the shear tensor field. Therefore, these two sets may be directly compared with each other. Fig~\ref{fig:filoverdensity_lambda} shows the probability density function of the logarithm of the normalised density, $\Delta=\rho/\rho_{\rm{mean}}$, for COWS filaments, V-web filament and all V-web cell. It is clear that the filaments lie along the ridges of the density field, in regions of higher density relative to the V-web filament cells. This is true independent of the value of $\lambda_{\rm th}$. The conclusion drawn from the apparent shift in the distributions, at a give $\lambda_{\rm th}$, is that the spine of the filaments are local density maxima in the two dimensional sections orthogonal to the local spine. Increasing $\lambda_{\rm th}$ shifts both distributions to higher densities, although the V-web filament cells are affected less than the COWS filaments. It is interesting that the width, or alternatively the height, of the COWS distributions are approximately constant. This suggests that a sample of filaments, independent of threshold, probes a constant variety of environments spanning approximately 2 orders of magnitude in density, where the threshold sets the average density of the sampled environment.

\subsubsection{Varying smoothing scale}
\label{sec:varysmoothing}

The second parameter that is investigated is the size of the Gaussian kernel used to smooth the CIC velocities for the V-web computation. This sets the effective resolution of the V-web. The top row of Fig.~\ref{fig:triplot_smoothscale} shows a V-web slice of the simulations for increasing kernel sizes, from left to right. Increasing the size of the kernel groups together cells into larger features, similar to zooming in. The relative sizes of the cosmic web types stay very similar but the absolute sizes of the features are increased. The number of filaments identified for each increasing size of the smoothing kernel is 39011, 14535, 5755 and 1560. The artefacts present in the skeleton are found to increase in size and do not appear to be significantly reduced in number as the kernel size is increased.

\begin{figure}
    \centering
    \includegraphics[width=\columnwidth]{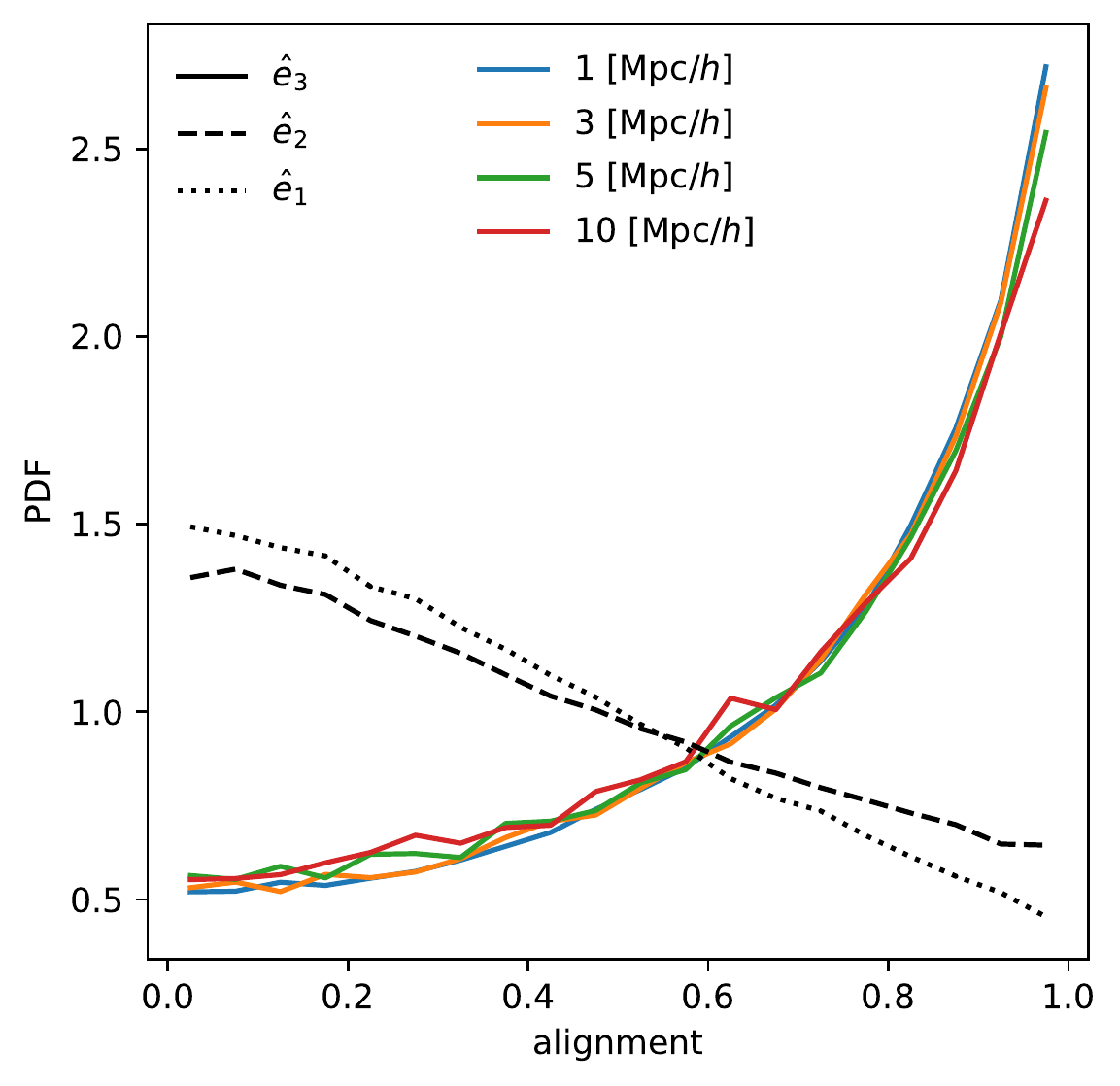}
    \caption{Probability density function (PDF) of the alignment between filament direction and the $\hat{e}_3$ eigenvector of the V-web. Filament direction is calculated using Equation~\ref{equ:fildirection}. Alignment is calculated as the absolute value of the dot product between the two vectors. Colours indicate different sizes of the Gaussian smoothing kernel. The V-web threshold, $\lambda_{\rm th}$, has been set to zero for all lines.}
    \label{fig:filalignment_smooth}
\end{figure}

Fig~\ref{fig:filalignment_smooth} shows the alignment of the filament orientation with the shear tensor eigenvectors but for changing values of the size of the Gaussian smoothing kernel. The alignment remains largely unaffected by the changes in the size of the smoothing kernel. The alignment displayed in Fig~\ref{fig:filalignment_smooth} is a sanity check that reaffirms the visual impression - the individual filaments identified with the algorithm under consideration are curvi-linear structures that are aligned along $\hat{e}_3$.

\begin{figure}
    \centering
    \includegraphics[width=\columnwidth]{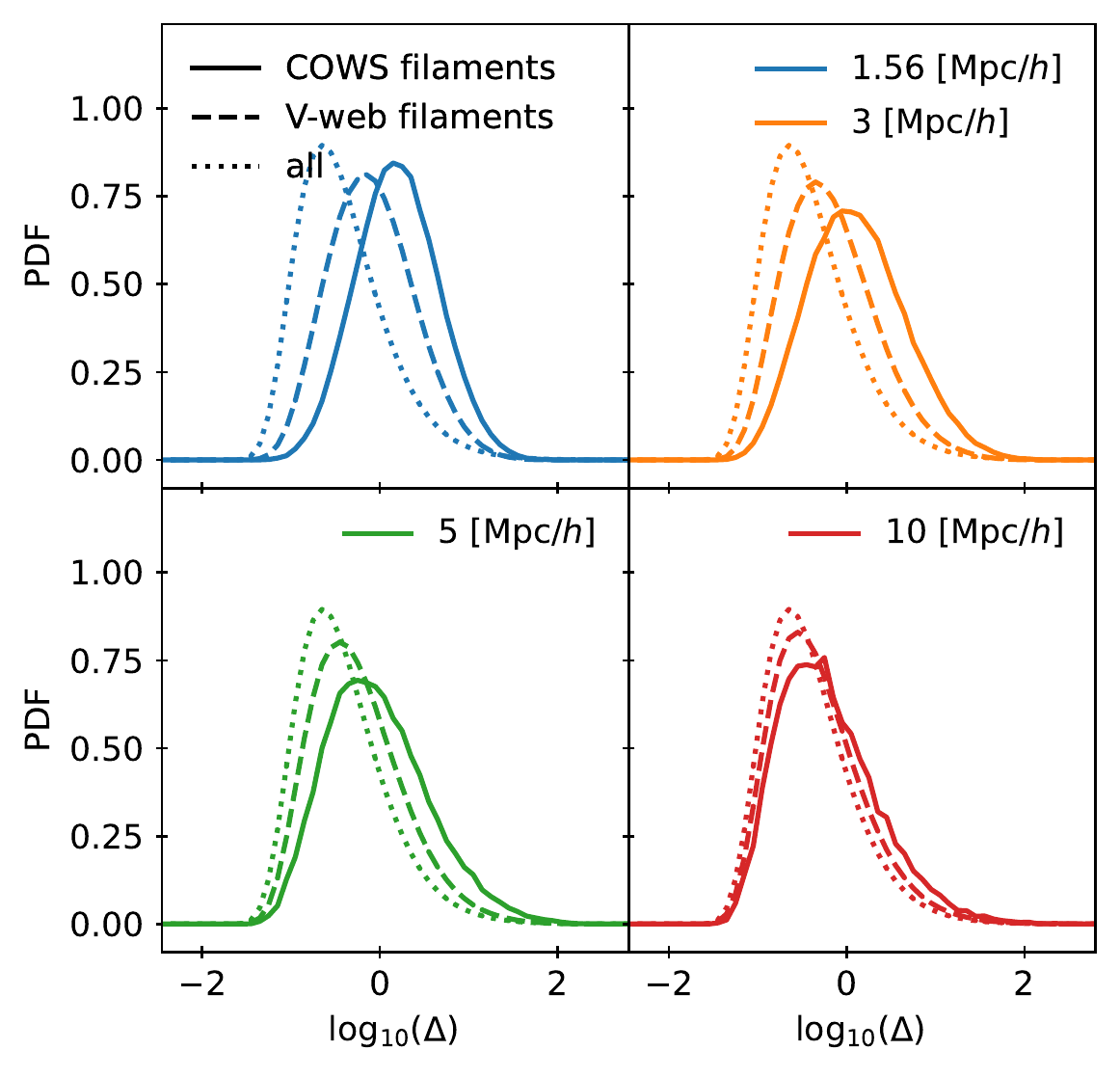}
    \caption{Probability density function (PDF) of the logarithmic normalised density for the identified filaments (solid), V-web filament cells (dashed) and all cells in the simulation (dotted). Colours indicate different sizes of the Gaussian smoothing kernel. The V-web threshold, $\lambda_{\rm th}$, has been set to zero for all lines.}
    \label{fig:filoverdensity_smooth}
\end{figure}

Fig~\ref{fig:filoverdensity_smooth} shows the probability density function of the logarithmic normalised density for varying sizes of the smoothing kernel. Increasing the size of the kernel shifts the distributions of V-web filament cells to lower density while the filaments tend to follow the parent V-web distribution more closely. For a kernel size of 10 Mpc/$h$, the density distribution in filaments very closely trace the V-web distribution. This is expected since large smoothing lengths homogenize the density and velocity field.

\subsection{Filament length distribution}
\label{sec:fillength}

\begin{figure}
    \centering
    \includegraphics[width=\columnwidth]{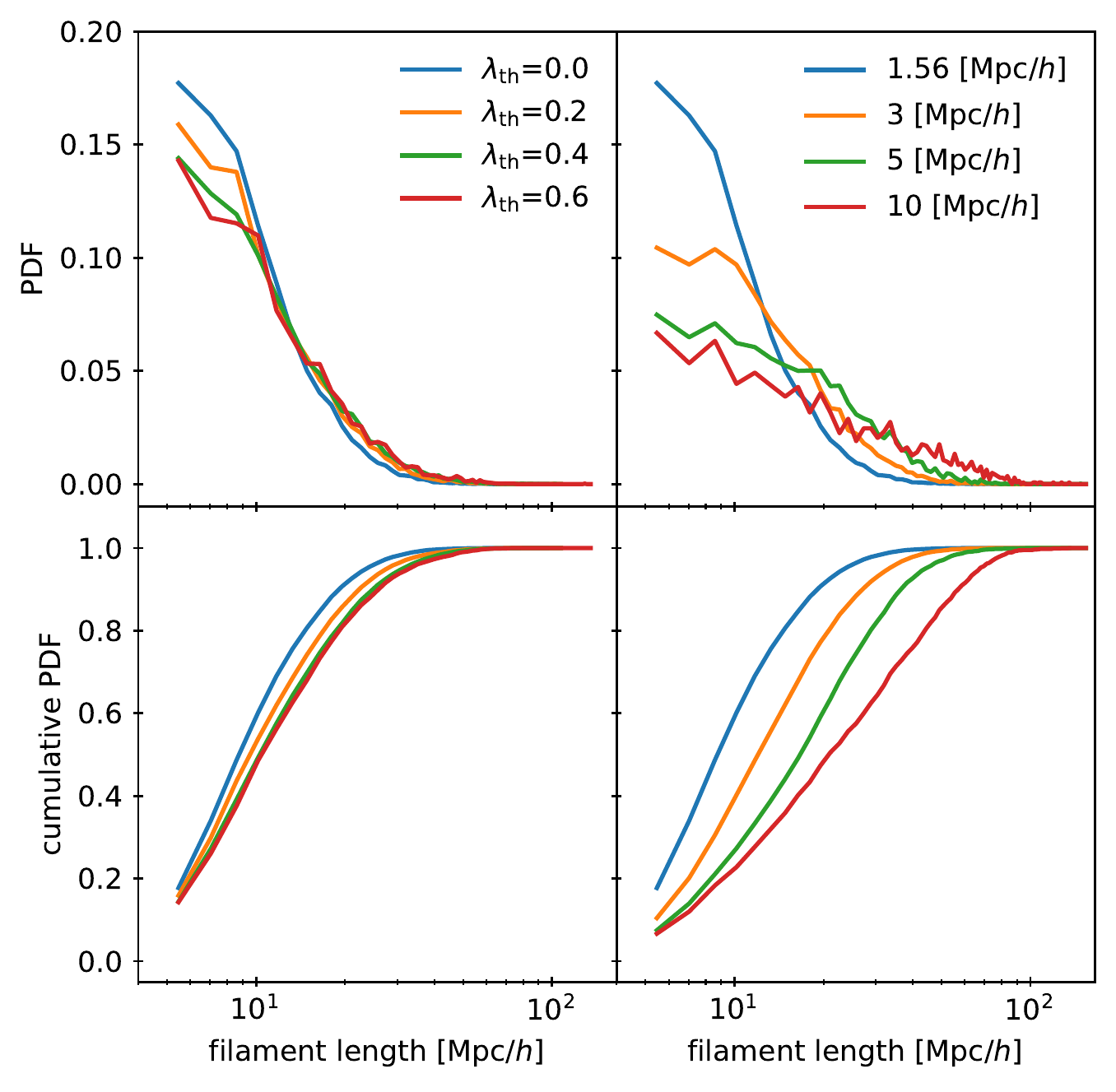}
    \caption{Probability density function (top) and the cumulative probability density function (bottom) of the filament length for different values of the threshold (left) and sizes of the Gaussian smoothing kernel (right).}
    \label{fig:fillengths}
\end{figure}

A selection of general filament properties are presented, the first of which is the distribution of filament lengths. Filament length is computed by summing the distances between neighbouring cell midpoints and adding one to account for the missing half cell at either end of the filament. Fig.~\ref{fig:fillengths} shows the probability density function of filaments for changing values of the threshold, $\lambda_{\rm th}$, (left) and the smoothing length (right). The filament population is dominated by the shortest filaments, independent of the parameter changes. Increasing $\lambda_{\rm th}$ returns longer filaments, although the change in the distributions is relatively small when compared to the effect  induced by varying the size of the smoothing kernel. Increasing the smoothing significantly reduces the amount of small filaments and changes the shape of the probability density function. This is because increasing the smoothing length pushes all cells towards lower densities, thereby smearing the filaments out.
Note that the mean filament length is related to the mean separation between junctions which is in turn correlated with the clustering of density peaks for a given cosmology. Therefore it is anticipated that the distribution of filament lengths could be used as a probe of the cosmological model. We differ this notion to future work. 

\subsection{Filament density profile}
\label{sec:fildensityprofile}

The radial density profile, in cylindrical coordinates, around the local spine of filaments is considered next. The  radial density profile at any position along the filament axis is computed by simply counting the number of particles within a given distance $r$ from the filament spine and within a given width (taken to be equivalent to one grid cell) along the filament spine, multiplied by the particle mass. The only free variable is the radial extent of the filament which is chosen to be $\approx$10 Mpc/$h$. Note that the radial distance of each particle is the shortest distance to the spine of the filament. The filament spine location can, at best, only be known to within half of a cell width because it is assumed that the filament spine runs through the centre of the filament cell. The minimum radius for the density profile is therefore set at $r\approx0.8$ Mpc/$h$ to remove the inner region where the central location of the filament spine is uncertain.

\begin{figure}
    \centering
    \includegraphics[width=\columnwidth]{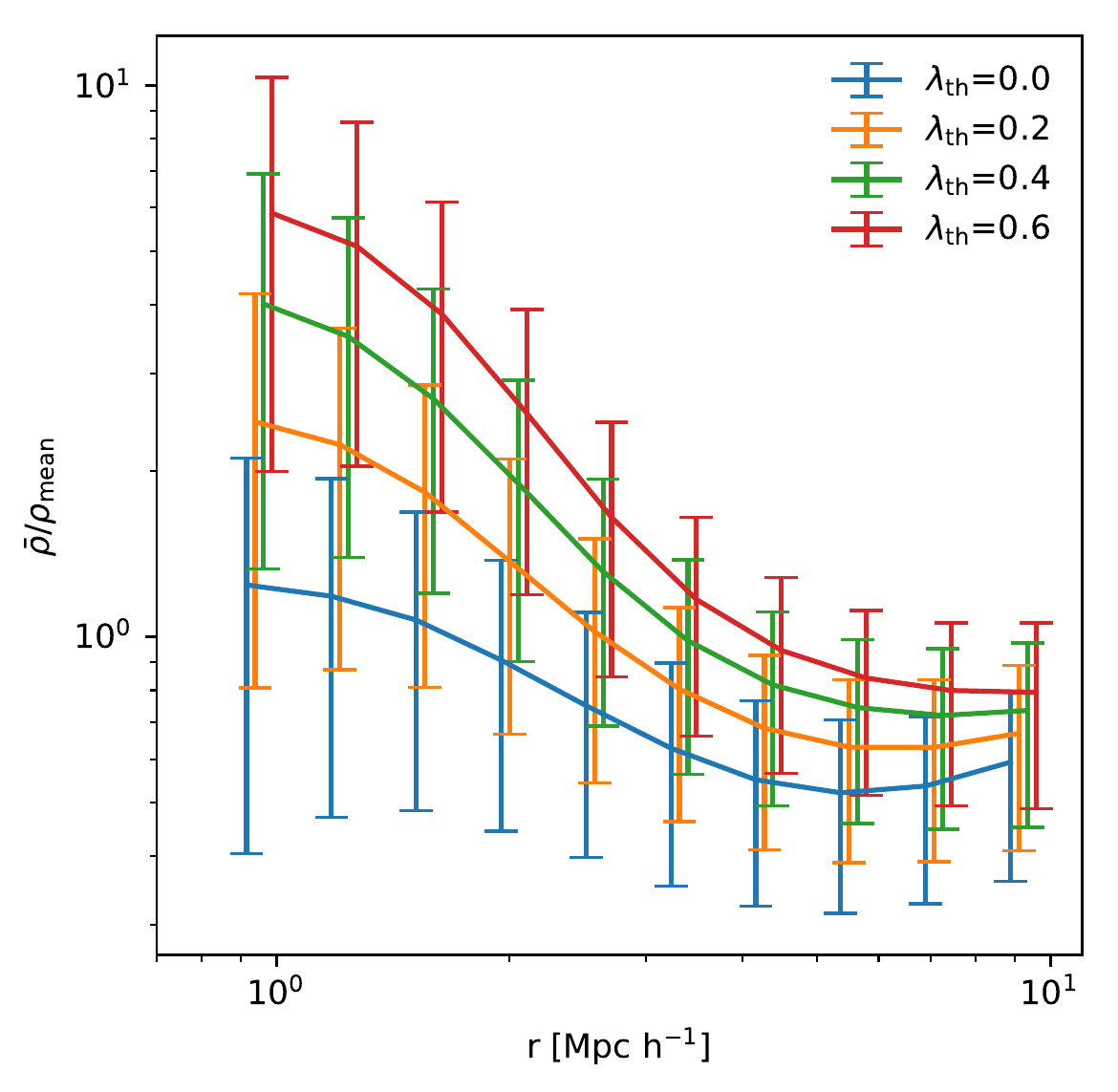}
    \caption{The median of the radial density profiles of all filament cells. Particles belonging to a filament cell are identified by considering a cylindrical region with length of one cell width centred at the cell midpoint. The radius of each particle is calculated as the shortest distance to the cylinder spine. The density is normalised by the mean global density of the simulation. The error bars show the 16th and 68th percentile of the distribution. Colours indicate different values of the V-web threshold. The lines have been shifted slightly along the x-axis for clarity. The true value corresponds to the $\lambda_{\rm th}=0$ line.
    }
    \label{fig:fildensityprofile}
\end{figure}

Fig.~\ref{fig:fildensityprofile} shows the median radial density profiles of all filament cells and the 16th and 68th percentile scatter for different thresholds. The density has been normalised by the mean density of the Universe. The density profiles constructed here are in general agreement with similar studies of filaments \citep[e.g.][]{cautun2014,galarraga2020}. Fig~\ref{fig:fildensityprofile} shows monotonically decreasing density profiles, confirming the visual impression that   the filaments' spines lie along density ridges. Increasing the threshold increases the amplitude of the density profile. The density profiles drop to $\bar{\rho}$ at increasing radii with increasing threshold. This is consistent with Fig~\ref{fig:filoverdensity_lambda} in that higher thresholds trace higher density regions, or alternatively, remove lower density regions from the V-web. There exists also a large amount of intrinsic scatter. This is to be expected since we are grouping together filaments that span, at least along the spine, two magnitudes in density irrespective of threshold (see Fig~\ref{fig:filoverdensity_lambda}). 

\section{Summary and Conclusion}
\label{sec:conclusion}
Filaments, multi-scale curvi-linear structures, are a generic feature of any density field that has evolved, under gravity, from an initial Gaussian random field of perturbations and form part of a greater whole called the Cosmic Web \citep{bond1996}. They are among the first ever features observed in the galaxy distribution (for example, the CfA ``stickman'' \citep{lapparent1986}), and more recently have been claimed to be critical environments in regulating galaxy formation in the context of galaxy spin \citep{tempel2013,wang2017,welker2020}, galaxy quenching (\eg \citealt{malavasi2021,kotecha2021}, however see also \citealt{liao2019,song2021}), and the intergalactic medium (\eg \citealt{klar2012,galarraga2021}).

Identifying filaments in the context of the Cosmic Web can be achieved using Hessian-based methods \citep{hahn2007,hoffman2012}. In this work, we focus on the V-web method, which classifies the Cosmic Web on a grid from the velocity field via the velocity shear tensor (see Equation~\ref{equ:vweb}). The V-web returns broad regions that are assigned to one of the four Cosmic Web types. It is however unclear how the identification of such regions can be re-organized into a catalog of filaments. Here a method is proposed to identify individual filaments from any peculiar velocity field, via the V-web, and applied to numerical simulation to study the properties of the resulting filaments (although, in principle, COWS can be applied to any regular grid-based Cosmic Web classifier).

The main aim of the investigated statistics is to test the resulting filaments and therefore validate the COWS method. It is difficult to evaluate the success of (any) filament finder since a detailed consensus on what defines a filament does not exist and many works that study filaments use different criteria to define them \citep{gonzalez2010,sousbie2011,tempel2013,cautun2014}. However, we investigate filament properties that we believe are common among many definitions and thus give a good indication of the ability to successfully detect cosmic filaments using our method. The main properties of the filament sample that are checked are whether their positions and orientations are as expected. The distribution of filament lengths and radial filament density profiles are also presented.

COWS filaments lie in regions of higher density relative to the average V-web filament environment. This can be understood because filaments tend to be densest along their spine, the axis of symmetry for a cylinder. The medial axis thinning algorithm returns the medial axis, the centre line, of the V-web filaments which closely trace the spine of the filaments, the density ridges that connect high-density region usually identified as knots.

Filament spines are also aligned with the main axis of expansion as characterized by the shear tensor. This geometric alignment is important as it indicates that the method proposed here is capturing the essence of the velocity field and its principal directions. The filaments we have identified also span a wide spectrum of lengths, the smallest being a few Mpc/$h$ (i.e. a couple of cells) while the largest are up to 100 Mpc/$h$. Their radial density profile can also be quantified and is in general qualitative agreement with other methods \citep[\eg][]{cautun2014,galarraga2020}), showing a monotonic decrease in density moving away from the filament spine.

The V-web has been employed on observational data for years because the cosmic velocity field can be inferred from accurate measurement of extra-galactic distances, redshifts and peculiar velocities as in the CosmicFlows projects \citep{tully2016,kourkchi2020}. This method is useful and powerful because a quantification of the peculiar velocity field can lead to the identification of basins of attraction and repulsion \citep{dubuy2019} as well as the discovery of super clusters like Vela \citep{kraan-korteweg2017,courtois2019} and Laniakea, our home supercluster \citep{tully2014}. This opens the door for COWS to be applied to observational data and verify filamentary structures in our local Universe, that before have only been identified through visual inspection, and perhaps uncover new, undiscovered cosmic filaments.

We believe that the utility of our method is significant. The ability to  identify individual V-web filaments directly from the velocity field will allow for the compilation of V-web filament catalogues and the study of individual objects. It will allow us to examine in detail the anatomy of individual filaments (besides their density profile, presented here) and to analyze matter flows and galaxy formation as a function of position in the filament. We expect to therefore use this method in the future to address many important science questions. Last but not least, this method can be applied to any field based cosmic web finder. We leave it to a future paper to compare and contrast the filaments produced by the T-web (based on the tidal shear tensor \citealt{hahn2007}) with those results presented here.

\section*{Acknowledgements}
This work has been done within the framework of the Constrained Local UniversE Simulations (CLUES) project. SP and NIL acknowledges financial support from the Deutsche Forschungs Gemeinschaft joint Polish-German research project  LI 2015/7-1. WH, MB, and KN acknowledge the support from the Polish National Science Center within research projects no. 2018/31/G/ST9/03388, 2020/39/B/ST9/03494. NIL acknowledge financial support from the Project IDEXLYON at the University of Lyon under the Investments for the Future Program (ANR-16-IDEX-0005). YH has been partially supported by the Israel Science Foundation grant ISF 1358/18. 

\section*{Data Availability}

The BAHAMAS simulation data used in this work are available upon reasonable request to SP.



\bibliographystyle{mnras}
\bibliography{main} 





\bsp	
\label{lastpage}
\end{document}